# Investigation and modeling of photocurrent collection process in multiple quantum well solar cells

Kasidit Toprasertpong*, Tomoyuki Inoue, Yoshiaki Nakano, and Masakazu Sugiyama

School of Engineering, the University of Tokyo, Bunkyo-ku, Tokyo 113-0032, Japan
***Corresponding author**, e-mail: toprasertpong@hotaka.t.u-tokyo.ac.jp

## Abstract

Solar cells employing quantum wells can enhance the light absorption but suffer from the difficulty in photocarrier extraction. Here, we analyzed the spectral response and the photocarrier collection mechanism of p-i-n multiple quantum well (MQW) solar cells using the effective-mobility model. Both the simulation and experimental results imply that the spatial profiles of electron and hole densities in MQWs play an important role in the carrier collection process. By considering the recombination increment under illumination, our findings suggest that the concept of the majority/minority carriers is important even in the intrinsic region: photogenerated electrons and holes only experience significant recombination when passing through the hole-rich and electron-rich regions, respectively. This can accurately explain the photocurrent behavior in cells with background doping, background illumination, and different MQW positions. Based on the experimental findings, we derived analytical formulae for carrier collection efficiency, which directly show the impact of each cell parameter and can be used for the systematic cell design.

*Keywords*: Multiple quantum wells, Modeling, Photocurrent, Carrier collection, Bias light, Device design

*Highlights*

- The behavior of photocurrent in low-mobility MQW solar cells is investigated.
- Carrier distribution in the i-region is an important factor for carrier collection.
- The impacts of background doping, bias light, and MQW position on EQE are clarified
- We derive analytical expressions for carrier collection efficiency.
- Our model is applicable to the cell design including well number and position.

# 1. Introduction

The implementation of nanostructures such as multiple quantum wells (MQW) in the photovoltaic application has been proposed as a promising approach for the efficient conversion of the sunlight spectrum into the electrical energy [1]. Such structures allow us to modify the optical properties in order to enhance the solar cell performance. A wide range of application of MQW solar cells, such as the photocurrent enhancement with small voltage loss [2], the restriction of radiative recombination [3], and the utilization of excitonic absorption [4], have been investigated. Another potential application which has already been commercially realized [5] is the bandgap engineering of subcells in multi-junction solar cells [3,6–9], where the bandgap combination is vital to the cell performance. MQW structures reduce the effective bandgap while keeping effectively lattice-matched with the substrate, offering alternative material choices for the bandgap optimization in multi-junction cells.

However, MQW absorbers have difficulty in carrier collection due to the obstruction of carrier transport by potential barriers [10]. Photogenerated carriers which are trapped and cannot escape out of the MQWs recombine and do not contribute to the photocurrent. MQWs are usually inserted in the intrinsic regions of the p-i-n junctions, where the internal electric field can enhance the carrier extraction [11]. Still, poor carrier collection has been observed in MQWs that have a large number of stacks and/or large band offsets [9,11–13].

To deal with this problem, there are a number of research works studying the mechanisms of carrier escape from the quantum wells (QWs). Escape models, such as thermal escape and tunnel escape, have been proposed, and detailed calculations of their escape rates have been discussed analytically [14,15]. However, a theoretical model which links the carrier escape (microscopic behavior) with the output photocurrent (macroscopic behavior) has not been well established. This limits the investigation of the device operation to the self-consistent numerical simulation [16], which is not straightforward to understand the device behavior. In addition, the numerical simulator is not always convenient to see the impact of cell parameters and optimize the cell design. Various analytical expressions have been employed to describe the collection efficiency of the charge carriers from MQWs such as $(1/\tau_{\mathrm{esc}})/(1/\tau_{\mathrm{esc}}+1/\tau_{\mathrm{rec}})$, $[(1/\tau_{\mathrm{esc}})/(1/\tau_{\mathrm{esc}}+1/\tau_{\mathrm{rec}})]^N$, and $(N/\tau_{\mathrm{esc}})/(N/\tau_{\mathrm{esc}}+1/\tau_{\mathrm{rec}})$, where $\tau_{\mathrm{esc}}$ is the carrier escape time from a QW to the next QW through a barrier, $\tau_{\mathrm{rec}}$ is the recombination lifetime, and $N$ is the stack number of QWs [12,13,17,18]. In addition to a variety of expressions, they are over-simplified. They do not take into account the backward movement of carriers, the driving force from the electric field, and the carrier distribution. Note that the above collection-efficiency expressions have only a weak dependence on the internal field[1], contradicting the fact that the internal field of the p-i-n structure is essential for the carrier collection from the MQW.

Many experimental studies have reported that the carrier collection from the quantum structures strongly depends on the surrounding conditions. It is well recognized that unintentional background doping extensively affects carrier collection [19–21]. This has been attributed to the weak electric field caused by band flattening, but only the band flattening cannot describe all the behaviors, which will be mentioned in this paper. The background illumination (bias light) has been reported to change the collection efficiency of the photoexcited carriers [22], but the mechanism behind this phenomenon has not been well-understood.

[1]For instance, the thermal escape rate from a QW with width $w$ and effective height $V_b$ under the electric field $E$ to the next well is proportional to $\exp\left[-\left(V_b-\frac{1}{2}wE\right)/k_BT\right]$ [14]. For $w$ = 5 nm, the escape rate is boosted by less than 10% under $E$ = 10 kV/cm compared to that under flat-band.

Here, we investigate the photocurrent collection mechanism, particularly the role of carrier distribution, in MQW solar cells. We employ the effective-mobility model, which takes into account the random carrier motion and the electric field, to simulate and analyze the results. Based on our experimental findings on various measurement conditions, we propose explicit expressions for carrier collection efficiency, which provides design rules for efficient photocarrier collection.

## 2. Experimental and simulation details

### *2.1. Sample preparation*

The p-i-n solar cells in this study were prepared by metal-organic vapor phase epitaxy with the growth condition explained in [21]. The thicknesses of the $2\times10^{18}$ cm$^{-3}$ doped GaAs p-emitter and the $1\times10^{17}$ cm$^{-3}$ doped GaAs n-base were both 200 nm. We applied a 25-nm InGaP window, but did not apply a back surface field and an anti-reflection coating. $In_{0.20}Ga_{0.80}As$ (5.4 nm)/$GaAs_{0.61}P_{0.39}$ (5.7 nm) strain-balanced MQWs were inserted in the i-regions, whose thickness was kept constant at 1100 nm for all samples. The atomic compositions and thicknesses of the MQWs were confirmed by X-ray diffraction (XRD) measurements.

We prepared two sets of samples as summarized in Table 1. The first set consisted of three MQW cells with different background doping levels in the i-regions. For convenience, we call them i-regions regardless of the background doping. The i-region of the p(p)n cell was unintentionally p-doped by carbon during the growth process, whereas the i-regions of the pin and p(n)n cells were additionally doped by small supply rates of $H_2S$. The background doping levels, shown in Table 1, were confirmed by Hall measurements in the GaAs test samples grown in the same condition. The second set consisted of cells with MQWs inserted in different positions in the i-regions: near the emitter (MQW-top), at the center (MQW-mid), and near the base (MQW-bottom). In this sample set, the compensation doping in the i-regions was carefully applied to guarantee the uniform electric field inside the MQWs. Samples within the same set were grown in the same batch and confirmed by XRD measurement to assure the identical structure of the MQWs.

**Table 1**

Sample details

| Sample | Well number | i-GaAs spacer [nm] | | Background doping [cm$^{-3}$][c] |
|---|---|---|---|---|
| | | Emitter side | Base side | |
| p(p)n | 45[a] | 300 | 300 | p-type $3\times10^{15}$ |
| pin | 45 | 300 | 300 | n-type $5\times10^{14}$ |
| p(n)n | 45 | 300 | 300 | n-type $8\times10^{15}$ |
| MQW-top | 27[b] | 100 | 700 | $< 5\times10^{14}$ |
| MQW-mid | 27 | 400 | 400 | $< 5\times10^{14}$ |
| MQW-bottom | 27 | 700 | 100 | $< 5\times10^{14}$ |

[a]The total MQW thickness was 500 nm.

[b]The total MQW thickness was 300 nm.

[c]Obtained from Hall measurements in GaAs test samples.

### *2.2. Carrier collection efficiency (CCE) measurement*

Since most photogenerated carriers can be collected from the MQW under sufficient reverse bias owing to high electric field, carrier collection efficiency (CCE) can be estimated by normalizing the current as

$$\mathrm{CCE}(V) = \Delta J(V) / \Delta J(V_{\mathrm{reverse}}), \tag{1}$$

where $\Delta J$ is the current increment after illumination, which corresponds to the *photocurrent*, $V$ is the voltage, $V_{\mathrm{reverse}}$ is the sufficiently high reverse-bias voltage [22]. For the illumination with monochromatic light having wavelength $\lambda$, Eq. (1) can also be written using external quantum efficiency (EQE) as

$$\mathrm{CCE}(V,\lambda) = \mathrm{EQE}(V,\lambda) / \mathrm{EQE}(V_{\mathrm{reverse}},\lambda). \tag{2}$$

CCE excludes the parasitic absorption (e.g. substrate absorption, which is included in the internal quantum efficiency measurement) and is straightforward in evaluating the collection process involving MQWs, but the careful interpretation is needed near the band edge, where the quantum-confined Stark effect arises.

EQE was measured using monochromatic light with a constant intensity of 2.5 mW/cm$^2$ passing through a chopper with a frequency of 85 Hz and captured by a lock-in amplifier, whose phase was carefully corrected. In the study of the light bias, additional AM1.5G light with an intensity of 100 mW/cm$^2$ (1 sun) was illuminated as a background without passing through the chopper. We focused on an operation voltage of 0.6 V, where the carrier transport degradation can be clearly observed while the effects from a voltage drop across the parasitic resistance [22] is sufficiently small. $V_{\mathrm{reverse}}$ was set to -6 V in the CCE measurement.

### *2.3. Simulation and effective-mobility model*

The experimental results were analyzed using the device simulator (PVcell, STR) [23]. The simulation was based on the conventional drift-diffusion model, solving the current continuity and Poisson's equations without additional complex physics. The Shockley-Read-Hall (SRH) recombination rate through deep-level defects formulated as

$$R_{\mathrm{SRH}} \approx \frac{np}{\tau_p n + \tau_n p} \tag{3}$$

was included in the current continuity equation. Here, $\tau_n$ and $\tau_p$ are the electron and hole SRH lifetimes, and $n$ and $p$ are the electron and hole densities. The illumination intensity, the applied voltage, the surface reflectance, and the device structure were set to be the same as the experimental condition except for the MQW structure.

To reduce the complexity of the MQW while maintaining its essence, we employed the effective-mobility model [24,25] in the device simulation. Instead of the periodic potential profile of wells and barriers, the MQW region is approximated as an equivalent bulk with a low effective mobility representing the sequential carrier trap-and-escape transport. We obtained the effective mobilities experimentally from the carrier time-of-flight measurement [26], which accidentally had similar values for electrons and holes [27]. The background doping level in the equivalent bulk was assumed to be approximately half of the value measured in GaAs, as the dopant ionization energy in wide-gap materials is relatively large [28] and the wide-gap GaAsP occupied approximately half of the MQW region. The other physical parameters of the equivalent bulk and their description are summarized in Table 2. The validation of the effective-mobility model is justified by a good agreement between the simulation results and the experimental ones. By using this model, all carrier dynamics in the MQW, including transport, generation, and recombination, can be treated in the similar way as the dynamics in a homogenous bulk.

**Table 2**

Physical parameters of the equivalent bulk representing MQW

| Parameter | Value |
|---|---|
| Electron effective mobility ($\mu_n$) | 0.28 cm$^2$/Vs |
| Hole effective mobility ($\mu_p$) | 0.28 cm$^2$/Vs |
| Bandgap | 1.29 eV [a] |
| Conduction band offset with GaAs ($\Delta E_C$) | 0.10 eV [a] |
| Valence band offset with GaAs ($\Delta E_V$) | 0.03 eV [a] |
| Effective conduction-band density of states ($N_{c,\mathrm{MQW}}$) | 2.2×10$^{18}$ cm$^{-3}$ [b] |
| Effective valence-band density of states ($N_{v,\mathrm{MQW}}$) | 4.6×10$^{17}$ cm$^{-3}$ [b] |
| Absorption coefficient ($\alpha$) | Measured by FTIR [c,d] (1.2×10$^4$ cm$^{-3}$ at 1.29 eV) |
| Radiative recombination coefficient ($B$) | 1.3×10$^{-9}$ cm$^3$/s [d] |
| SRH lifetimes ($\tau_n$, $\tau_p$) | Fitting parameters [e] |

[a]Obtained from the photoluminescence peak at room temperature and the estimated positions of the 1e and 1hh energy levels.

[b]Ref [24,25]

[c]Fourier transform infrared spectroscopy

[d]$B$ was obtained from the absorption coefficient $\alpha$ through the van Roosbroeck-Shockley relation and the photon reabsorption probability [24,25]. Since the measured absorption coefficient was used, the actual oscillator strength in the MQWs has been included in the equivalent bulk.

[e]$\tau_n$ = 140 ns and $\tau_p$ = 140 ns for the background-doping sample set and $\tau_n$ = 300 ns and $\tau_p$ = 35 ns for the MQW-position sample set. The different lifetime values between the two sample sets were due to the different growth batches.

## 3. Experimental results and discussion

### 3.1. *Effect of background doping*

Fig. 1 shows the EQE of the p(p)n, pin, and p(n)n MQW solar cells measured at -6 V and 0.6 V. All the MQW cells have almost the same EQE at the high reverse bias condition regardless of the background doping. At high reverse bias, EQE simply reflects the device absorptivity, which is the same for cells containing the identical MQW structure. On the other hand, applying forward bias significantly degrades EQE due to the poor carrier collection at the low field. In this way, the ratio of these EQEs in Eq. (2) accurately extracts the information on carrier collection at a forward bias.

Fig. 2(a) shows the CCE at 0.6 V obtained from the ratio of these EQEs. The p(p)n and p(n)n cells suffer from poor collection of carriers excited by short- and long-wavelength light, respectively, whereas the pin cell can efficiently collect carriers from a wide wavelength range. In most research so far, the poor carrier collection has been explained by the background doping flattening the band, as depicted in Fig. 2(c). While the i-region of the pin cell has the uniform field, the field region covered less than half of the MQW region in the p(p)n and p(n)n cells, making it difficult to sweep carriers out of the MQWs. This band flattening effect can well explain the poor carrier collection at the particular illumination condition, but not the excitation-wavelength dependence. In particular, carriers from short-wavelength excitation can be efficiently collected in the p(n)n sample despite the fact that the band is almost flat in the MQW region.

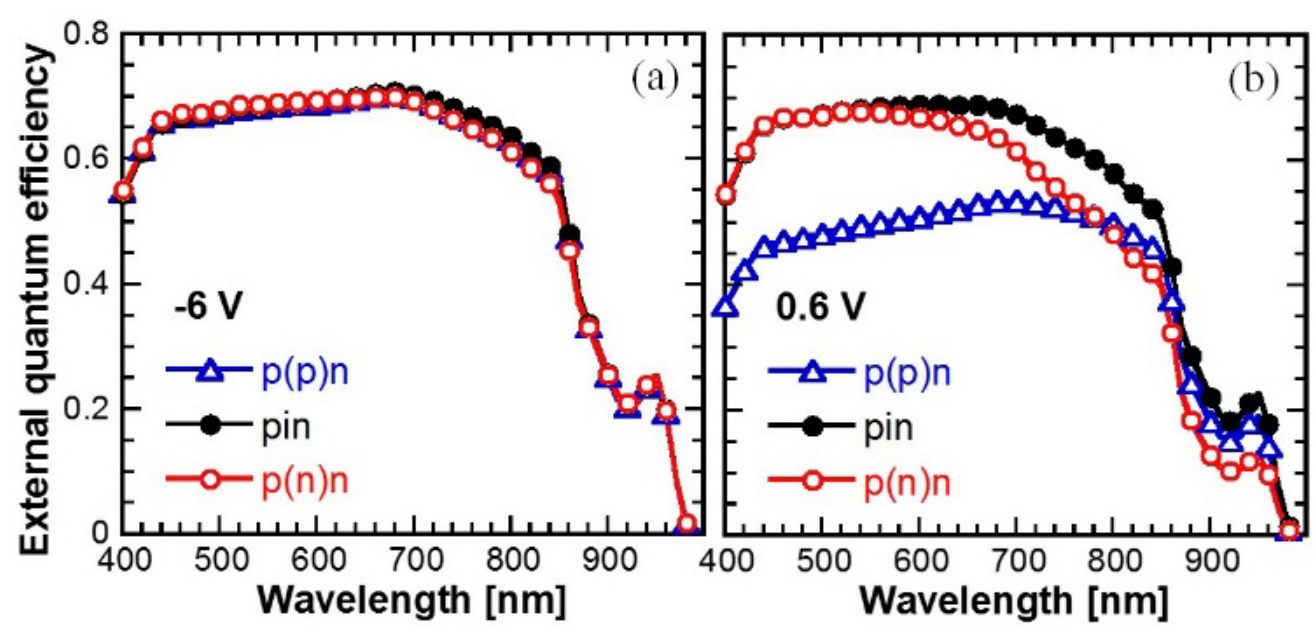

**Fig. 1.** EQE spectra of MQW cells with different background doping levels measured at (a) -6 V and (b) 0.6 V. No bias light was applied.

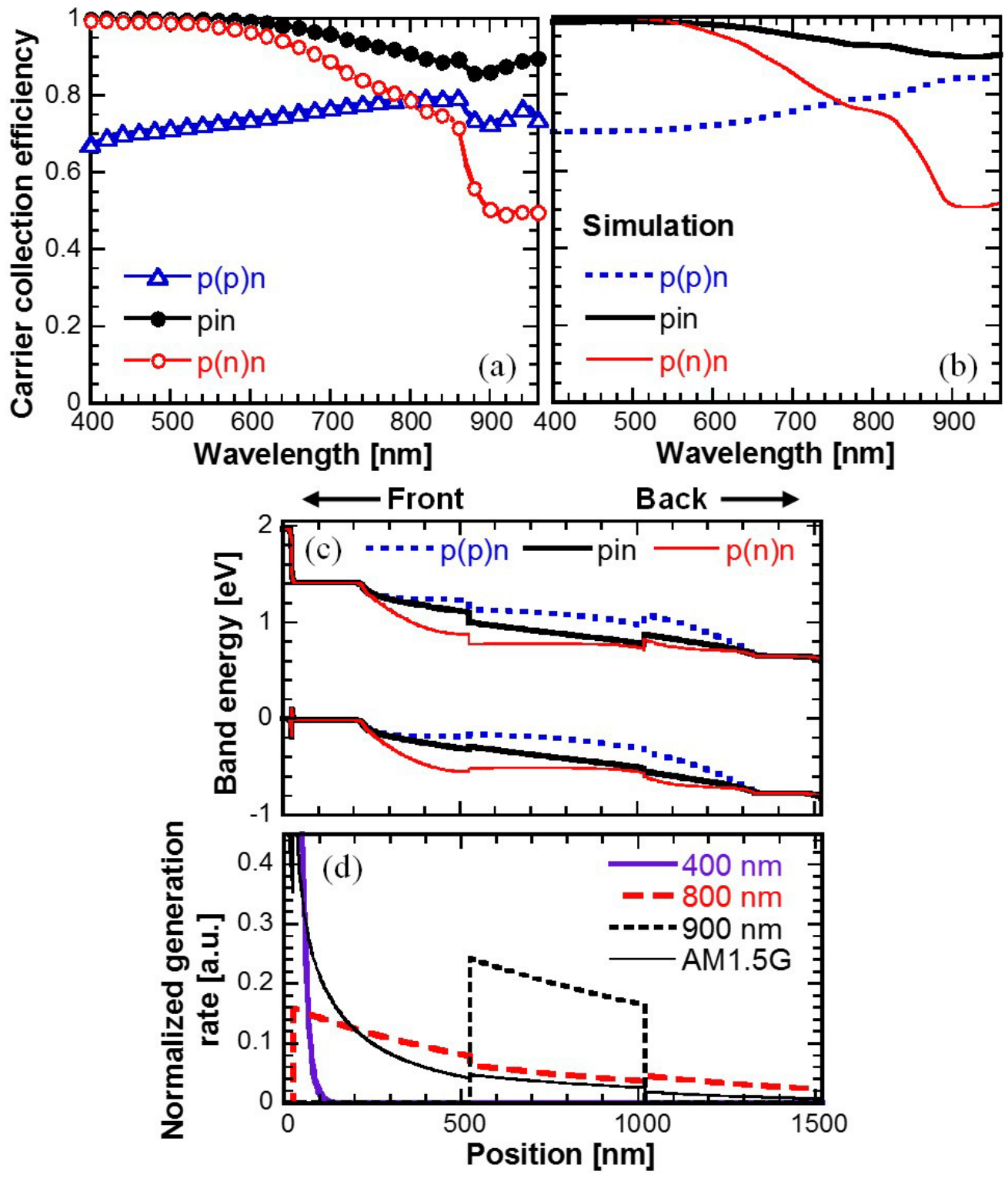

**Fig. 2.** (a) Measured wavelength-dependent CCE at 0.6 V for different background doping levels. Simulated (b) CCE and (c) band diagrams at 0.6 V using the effective-mobility model, in which the MQW is replaced with an equivalent bulk. (d) Normalized photogeneration rates under 400-nm-wavelength, 800-nm-wavelength, 900-nm-wavelength, and AM1.5G illumination.

In order to understand this behavior, we simulated the operation of these cells using the model explained in Section 2.3. We focused on the simulation of the CCE, obtained from Eq. (1) and the simulated photocurrent, rather than the EQE in order to avoid the complicated estimation of parasitic absorption and to discuss the collection process directly. The simulated CCE is shown in Fig. 2(b). The SRH lifetimes of electrons $\tau_n$ and holes $\tau_p$ in the equivalent bulk representing the MQWs are the fitting parameters that reproduce the measured CCE of the pin cell and were found to be both 140 ns, which are close to the lifetimes estimated from the electroluminescence measurement in similar MQW structures (~100 ns) [29]. By only varying the background doping level while keeping other parameters the same, the simulated CCE of the p(p)n and p(n)n samples in Fig. 2(b) can adequately reproduce the tendency of the experimental data in Fig. 2(a). This implies that, in spite of many assumptions and approximations included, the effective-mobility model keeps the essence of MQW and can sufficiently describe the carrier collection process inside the MQW. In this way, understanding the physics behind these simulation results will provide useful information on the photocurrent collection mechanism in actual devices.

Firstly, we discuss the behavior under the 400-nm illumination. This wavelength has a short absorption length and excites carriers only in GaAs above the MQW as depicted in Fig. 2(d). Generated holes are easily collected at the p-emitter whereas generated electrons have to travel through the MQW toward the n-base. Thus, the photocurrent entering the MQW consists of only the electron component, making the analysis simple. Fig. 3 shows the simulation results of the carrier densities, the recombination rates, and the electron photocurrent in the i-region at 0.6 V. Here, instead of the total recombination rate and the total current, we consider the *increment* after illumination. The recombination increment due to illumination directly corresponds to the recombination of the photogenerated carriers, i.e. the failure of photocurrent collection, and the current increment corresponds to the photocurrent. In this way, the behavior of the dark current can be excluded, allowing us to focus on the behavior of the photocurrent. The increment of recombination in Figs. 3(d)-(f), which is found to be dominated by the SRH process as can be expected from the characteristics of the i-regions, is large in the entire MQW region of the p(p)n sample and in the hole-rich ($p > n$) region of the pin sample. This causes a drop of electron photocurrent flowing through the MQW [Figs. 3(g)-(h)]. On the other hand, the recombination increment is small in the p(n)n sample and in the electron-rich ($n > p$) region of the pin sample, resulting in comparatively small electron photocurrent drop in these regions [Figs. 3(h)-(i)]. This corresponds to the low and high short-wavelength CCE in the p(p)n and p(n)n sample, respectively.

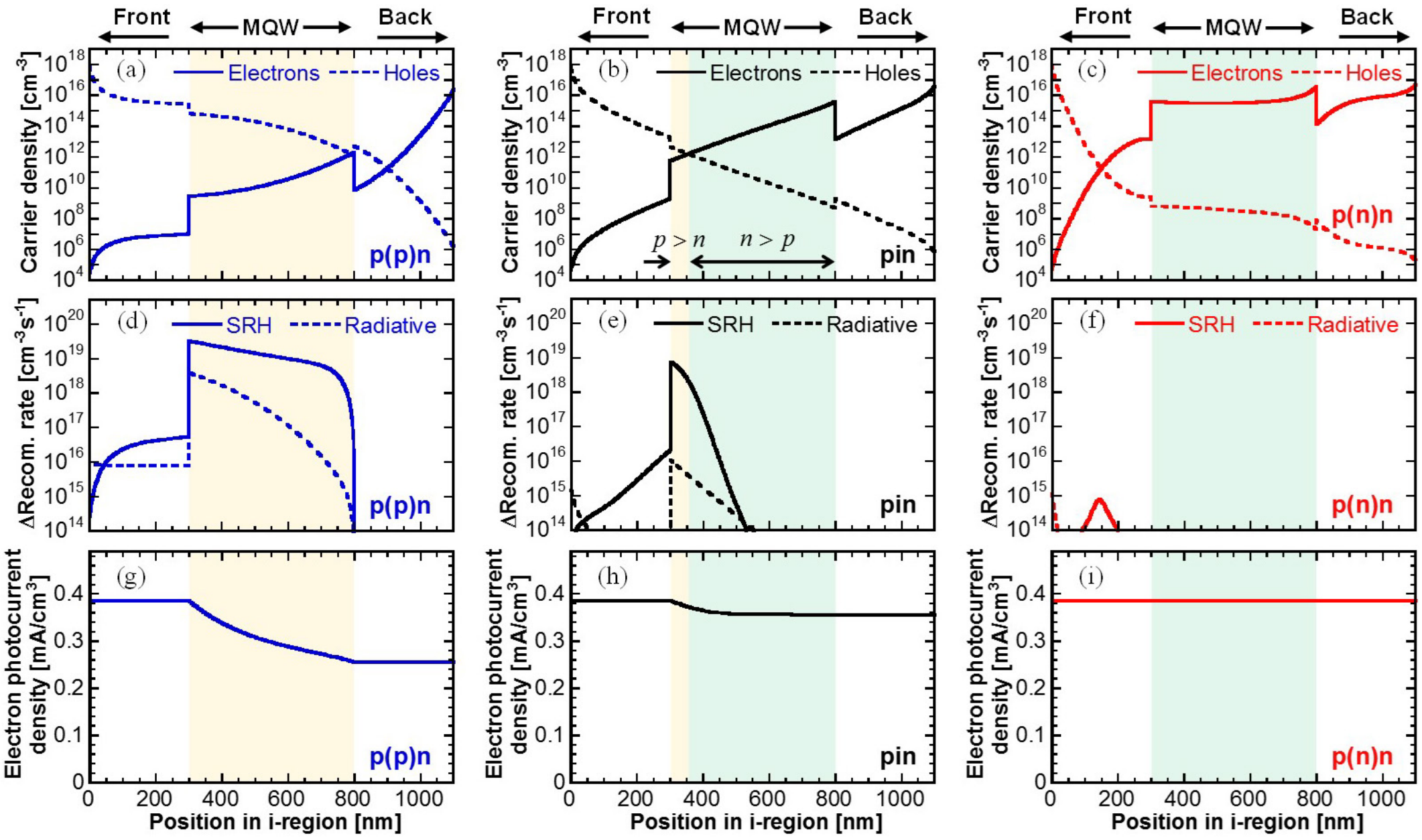

**Fig. 3.** Simulation results of the p(p)n, pin, and p(n)n MQW cells at 0.6 V. (a)-(c) Electron and hole densities in the dark condition. (d)-(f) Increment of SRH and radiative recombination rates under 400-nm light. (g)-(i) Electron photocurrent densities, obtained from the difference between the electron current densities without and with the 400-nm-wavelength illumination. Only the i-regions are shown. The shaded area indicates the MQW regions with electron-rich and hole-rich profiles.

This behavior can be explained by the following: when electrons flow in the MQW, the electron density changes by $\Delta n$ and the SRH recombination in Eq. (3) increases by

$$\Delta R_{\mathrm{SRH}} \approx \frac{\tau_n p^2 \Delta n}{\left(\tau_p n + \tau_n p\right)^2} \approx \begin{cases} \Delta n / \tau_n & ; p > n \\ 0 & ; n > p \end{cases} . \tag{4}$$

Eq. (4) implies that the recombination of photogenerated electrons and, consequently, the electron photocurrent drop mainly occur in the hole-rich region. In the p(n)n MQW cell, we can conclude from the above findings that the suppression of the recombination process overcomes the poor carrier transport even though the MQW is almost outside the field region, allowing photogenerated electrons to be efficiently collected and the short-wavelength CCE to be almost 1. The low short-wavelength CCE in the p(p)n cell is due to both the wide hole-rich region, which boosts the carrier recombination, and the low field due to the band flattening, which degrades the carrier transport.

For longer excitation wavelengths (500 nm < $\lambda$ < 870 nm), which have smaller absorption coefficients, a larger portion of carriers are generated near the n-base as shown in Fig. 2(d). Electrons photogenerated below the MQW are efficiently collected at the n-base while holes flow toward the MQW. The p(n)n cell, whose MQW region is electron-rich, begins to suffer from poor collection of the hole photocurrent, resulting in the degradation of long-wavelength CCE. The CCE of the p(p)n cell, on the other hand, has a slight improvement at long wavelengths owing to the smaller fraction of carriers generated above the MQW, thus smaller electron photocurrent flowing through the MQW.

The above analysis implies that the concept of the *majority/minority carrier region* is important in the carrier collection process even when the MQW is in the i-region. That is, the carrier collection efficiency is determined by the transport distance and the transport properties, e.g. field profile, *only* in the corresponding minority-carrier region (hole-rich region for photogenerated electrons and electron-rich region for photogenerated holes). Note that the carrier distribution in the i-region, unlike the n- or p-region, originates from the quasi-thermal equilibrium condition with the neighboring doped regions. In the following sub-sections, we apply the similar analysis to get insight into the carrier collection behavior in cells with bias light and with different MQW positions.

## 3.2. *Effect of bias light*

Fig. 4 shows the EQE and CCE of the p(p)n, pin, p(n)n MQW solar cells measured at 0.6 V under a 1-sun bias light. By illuminating the background light during the EQE measurement, only the p(p)n cell shows a significant improvement from the poor carrier collection efficiency under no bias light [Figs. 1(b), 2(a)].

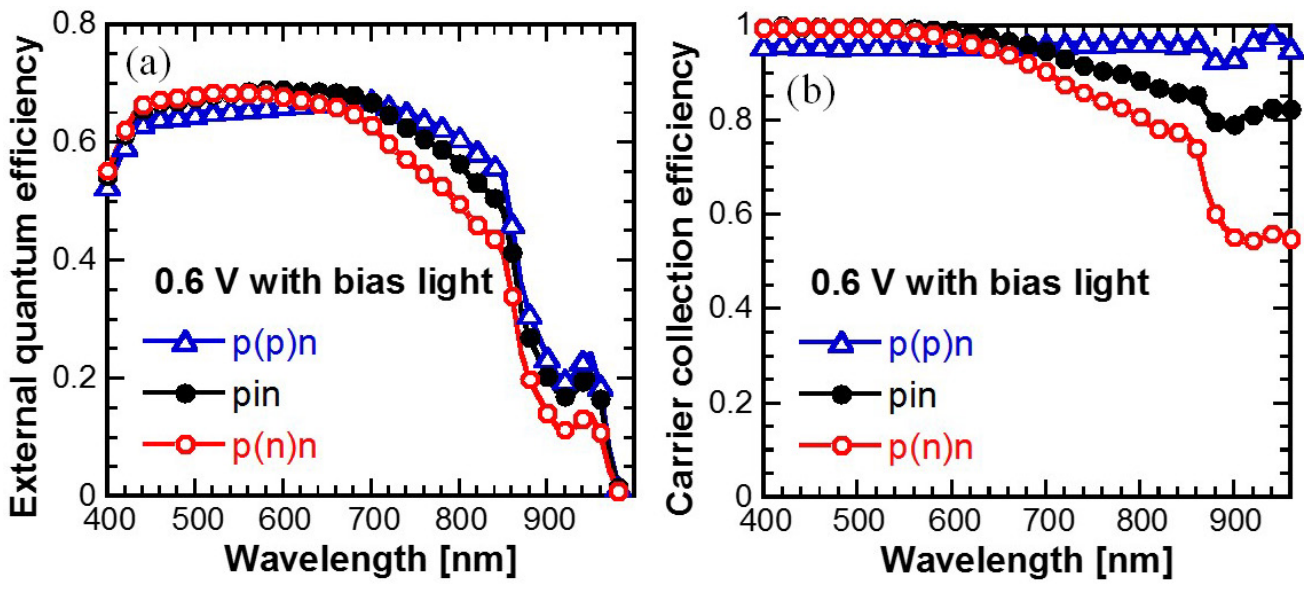

**Fig. 4.** (a) EQE and (b) CCE spectra of MQW cells with different background doping levels at 0.6 V under 1-sun AM1.5G bias light.

As can be confirmed from the band simulation in Fig. 5 that the band profile is almost unchanged under 1-sun bias light, the significant enhancement in carrier collection is not attributed to the change of the field profile. To understand this behavior, we firstly investigated the increment of the carrier densities, recombination rates, and electron photocurrent in the p(p)n MQW cell for the 400-nm-light response, where only photogenerated electrons flow through the MQW. The simulation results are shown in Fig. 6. Here, the increments of recombination rates and electron current are defined by the difference between the values under AM1.5G + 400 nm illumination and the values under AM1.5G illumination. As can be seen from Fig. 6(a), the AM1.5G illumination remarkably changed the carrier distribution profile of the p(p)n cell: the hole-rich region width decreases from the entire MQW [Fig. 3(a)] to only a part of the MQW region. This is because of the low effective mobility in the MQW that easily enhances the carrier accumulation under illumination even for non-concentrated light. The recombination increment in Fig. 6(b) is suppressed in the electron-rich region, resulting in the negligible drop of the electron photocurrent in this region as depicted in Fig. 6(c) and the recovery of the short-wavelength CCE.

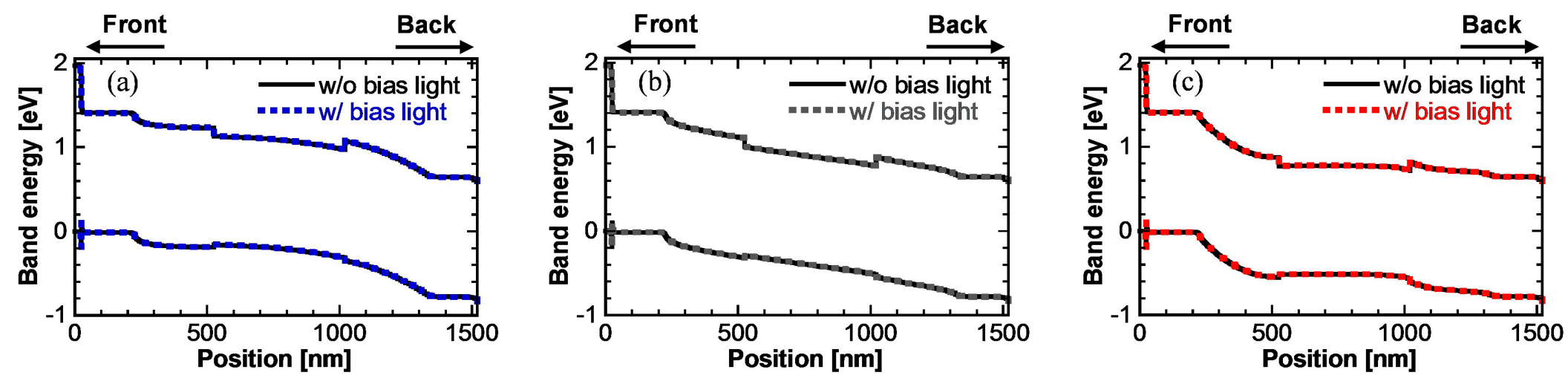

**Fig. 5.** Band diagrams at 0.6 V without and with 1-sun AM1.5G bias light for the (a) p(p)n, (b) pin, and (c) p(n)n MQW solar cells.

For long-wavelength excitation (500 nm < $\lambda$ < 870 nm), the CCE of the p(p)n cell recovers under bias light similarly to the short-wavelength excitation. As can be seen in Fig. 2(d), the fraction of electron photocurrent flowing through the MQW (electrons generated above the MQW and flowing toward the n-base) is larger than the hole photocurrent (holes generated below the MQW and flowing toward the p-emitter) due to the exponential profile of light intensity. Therefore, the decrease of the hole-rich region and, consequently, the enhancement in the electron collection boost CCE as well at long wavelengths.

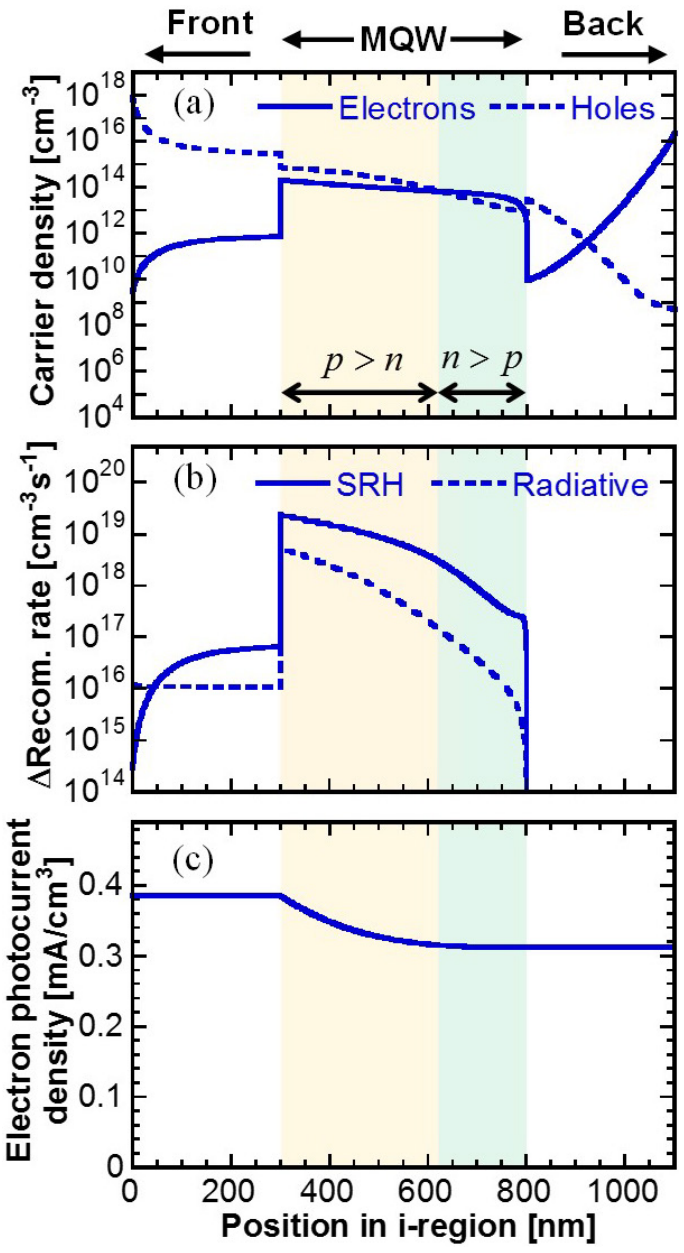

**Fig. 6.** Simulation results in the i-regions of the p(p)n MQW cells at 0.6 V under AM1.5G 1-sun bias light. (a) Electron and hole densities under bias light. (b) Increment of recombination rates under the 400-nm additional illumination. (c) Electron photocurrent density generated by the 400-nm light. The shaded areas indicate the electron-rich and hole-rich regions inside the MQW.

By contrast, the effect of bias light is comparatively small in the pin and p(n)n cells. The AM1.5G bias light accumulates electrons in the MQW rather than holes due to the larger fraction of carriers generated by AM1.5G near the p-emitter than the n-base [Fig. 2(d)], resulting in a larger electron current flowing into the MQW. This electron accumulation, however, has only a small effect on these two cells since they already have wide electron-rich regions before applying bias light [Figs. 3(b)-(c)].

This finding suggests that the quantum efficiency measurement should be conducted under bias light to ensure the same carrier distribution profile as the actual device operation.

### 3.3. *Effect of MQW position*

In this part, we discuss the effect of the MQW position inside the i-region using the MQW-top, MQW-mid, and MQW-bottom described in Table 1. Figs. 7(a)-(b) show their EQE and CCE at 0.6 V under 1-sun bias light. Despite the identical MQW and the uniform electric field owing to the negligible background doping, the carrier collection is different among samples whose differences are only in the position of the MQW.

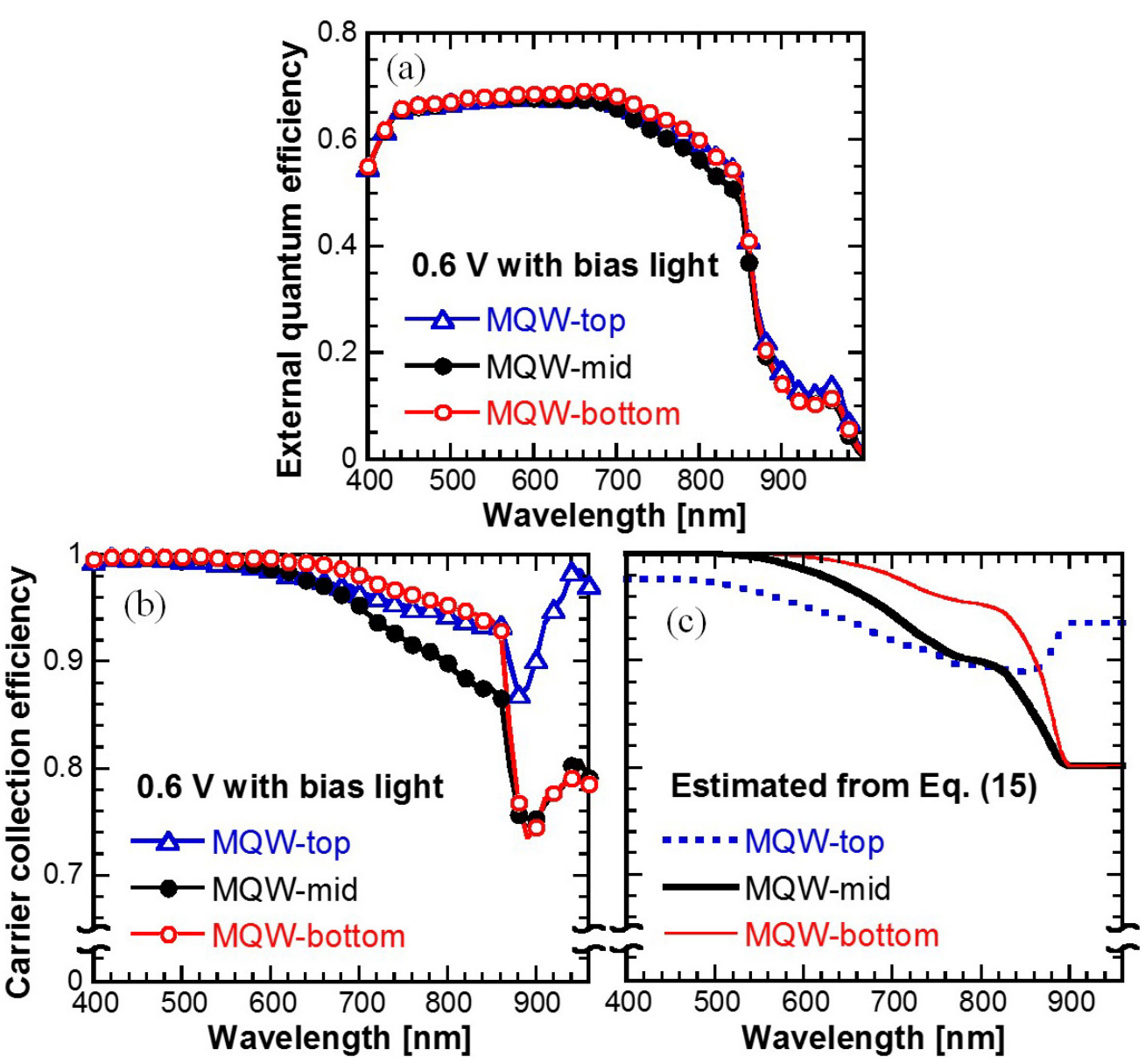

**Fig. 7.** (a) EQE and (b) CCE spectra of cells with the MQW inserted at the top, middle, and bottom of the i-region at 0.6 V under 1-sun bias light. (c) CCE estimated using Eq. (15).

We, again, investigated by simulating the carrier distribution profile as shown in Figs. 8(a)-(c). It can be seen that the MQW-mid and MQW-bottom have the electron-rich region in the entire MQW region, whereas MQW-top partly has the hole-rich region inside the MQW. This is due to the electron accumulation from the AM1.5G bias light. Note that if there is no bias light, relatively symmetric carrier distribution can be found in the MQW-mid. For short-wavelength illumination, as previously explained, there is only electron photocurrent in the MQW coming from the front part of the samples, and it experiences no difficulty in flowing through the MQW due to the wide electron-rich region, with only a slight (< 0.5%) CCE drop in MQW-top.

For longer wavelengths (500 nm $< \lambda <$ 870 nm), CCEs of the MQW-top, MQW-mid, and MQW-bottom start to drop to below 0.995 at wavelengths of 500, 560, and 620 nm. The GaAs thicknesses for a 98% absorption at these threshold wavelengths are 365, 616, and 912 nm [30], corresponding to the GaAs thicknesses of 300, 600 and 900 nm, respectively, above the MQWs (200-nm p-emitter + top i-spacer). Light with wavelengths longer than the threshold allows carriers to be photoexcited beyond the MQW and generates the hole photocurrent which cannot be efficiently collected when crossing the electron-rich regions. Under the 800-nm illumination, the MQW-mid was found to have the poorest carrier collection. This is due to the fact that in the MQW-mid, more light penetrates the MQW and generate a larger hole photocurrent than the MQW-bottom, while the electron-rich region is wider than that in the MQW-top.

For direct excitation in the MQW ($\lambda$ > 870 nm), the MQW-top shows a remarkably higher CCE than the other two samples. These wavelengths generate carriers in the entire MQW region and are not absorbed outside the MQW. The collection of carriers generated in the lower (upper) part of the MQW is limited by the hole (electron) transport in the electron-rich (hole-rich) region toward the p-emitter (n-base), whereas the majority carriers in each region can be efficiently extracted out of the MQW. As illustrated in Figs. 8(d)-(e), the average transport length in these regions is the shortest when the carrier distribution is symmetric. This is the reason why the MQW-top, whose carrier distribution in the MQW is comparatively balanced, has a high CCE for the direct photogeneration. We will discuss this behavior mathematically in the next section.

Note that the quantum-confined Stark effect shifts the absorption wavelength at the reverse bias and makes the CCE calculation using Eq. (2) overestimated or underestimated, depending on the wavelengths, near the absorption edge. This appears as a peak-shape artifact in the measured CCE spectra.

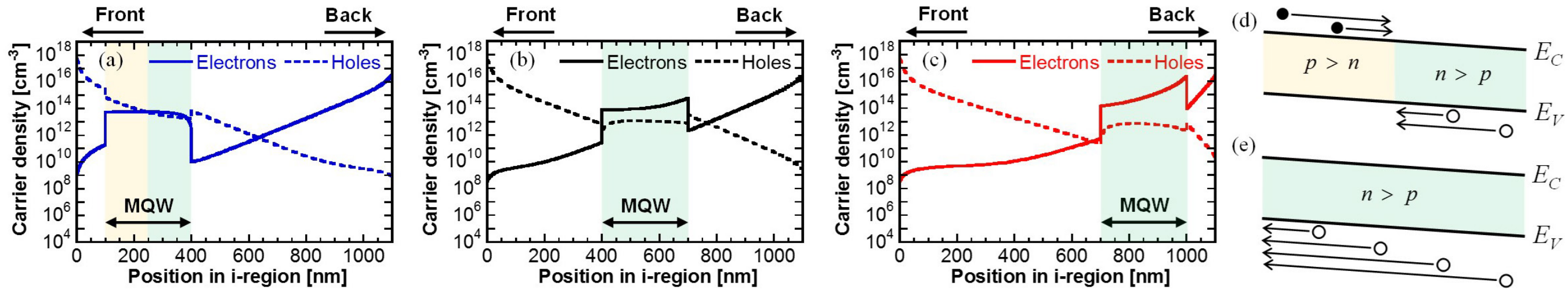

**Fig. 8.** Simulated electron and hole densities in the i-regions of MQW cells when the MQW is inserted (a) near the p-emitter (MQW-top), (b) at the center (MQW-mid), and (c) near the n-base (MQW-bottom) at 0.6 V under 1-sun bias light. The shaded areas indicate the MQW regions with different carrier distribution. Transport distance of the minority carriers directly photogenerated inside the MQW when the electron-rich region covers (d) a half and (e) the entire MQW region. The average transport length of the minority carriers is the shortest when the carrier distribution is symmetric.

## 4. Formulae

In the last section, the carrier collection behavior of MQW p-i-n solar cells is investigated experimentally and analyzed in a qualitative manner. In this section, the underlying physics is analyzed mathematically based on the above findings and expressed in a set of simple formulae. To simplify the analysis, the main assumptions here are that

- The increments of carrier density, recombination, and current under illumination correspond to the dynamics of photogenerated carriers
- The drift process dominates the transport of photogenerated carriers in the i-region
- The SRH process dominates the recombination increment as discussed in the last section

### 4.1. Collection of carriers crossing MQW

The increase of electron density $\Delta n$ in the MQW due to the flow of electron photocurrent $\Delta J_n$ is determined by the relation

$$\Delta J_n = q\mu_n E\Delta n \ , \tag{5}$$

where $q$ is the elementary charge, $\mu_n$ is the electron effective mobility in the MQW, and $E$ is the electric field. The SRH recombination increment at depth $x$ causes the electron photocurrent drop by

$$\frac{d}{dx}\left(\Delta J_n\right) = -q\Delta R_{\mathrm{SRH}} = \begin{cases} -\Delta J_n / \mu_n \tau_n E & ; p > n \\ 0 & ; n > p \end{cases} , \tag{6}$$

where the relation among $\Delta R_{\mathrm{SRH}}$, $\Delta n$ and $\Delta J_n$ in Eqs. (4)-(5) is used.

The solution for $\Delta J_n$ in Eq. (6) is an exponential function, and the electron collection efficiency $\mathrm{CCE}_n$, obtained from the ratio of the electron photocurrent extracted from MQW ($\Delta J_{n,\mathrm{out}}$) to the electron photocurrent entering the MQW ($\Delta J_{n,\mathrm{in}}$), is expressed by

$$\begin{aligned} \mathrm{CCE}_n &= \frac{\Delta J_{n,\mathrm{out}}}{\Delta J_{n,\mathrm{in}}} \\ &= \exp\left[ -\int_{p>n} \frac{dx}{\mu_n \tau_n E} \right] = \exp\left[ -\left\langle \frac{1}{\mu_n \tau_n E} \right\rangle L_{p>n} \right], \end{aligned} \tag{7}$$

where $L_{p>n}$ is the width of the hole-rich region in the MQW. This explains the carrier collection behavior in the p(n)n and p(p)n cells: electrons can be efficiently collected in the p(n)n cell even at a low field strength $E$ thanks to $L_{p>n}$ being close to zero, whereas they strongly recombine in the p(p)n cell, which has both large $L_{p>n}$ and low $E$. Although only three different background doping levels were prepared in this study, the collection behavior under other doping levels can be predicted using Eq. (7). With higher p-type background doping, $\mathrm{CCE}_n$ in the p(p)n cell is expected to be further degraded due to the lower $E$ caused by the band flattening effect as well as the larger $L_{p>n}$. On the other hand, for the n-type background doping, the $\mathrm{CCE}_n$ in the p(n)n cell is expected to remain close to 1 regardless of the doping level as long as $L_{p>n} \approx 0$.

For MQW with uniform effective mobility, lifetime, and field, electron collection efficiency $\mathrm{CCE}_n$ and hole collection efficiency $\mathrm{CCE}_p$ across the MQW can be written by a similar expression using the effective mobility $\mu_{\mathrm{MQW}}$ and the lifetime $\tau_{\mathrm{rec}}$ of the minority carriers:

$$\mathrm{CCE}_{n \text{ or } p} = \exp\left[ -\frac{L_{\mathrm{minor}}}{\mu_{\mathrm{MQW}} \tau_{\mathrm{rec}} E} \right] = \exp\left[ -\frac{L_{\mathrm{minor}}}{L_d} \right], \tag{8}$$

where $L_{\mathrm{minor}}$ is the minority-carrier region width and

$$L_d = \mu_{\mathrm{MQW}} \tau_{\mathrm{rec}} E \tag{9}$$

is the minority-carrier drift length.

## *4.2. Collection of carriers generated inside MQW*

We consider the collection of carriers directly generated in the MQW with a total thickness of $L_{\mathrm{MQW}}$, uniform electron and hole drift lengths of $L_{d,n}$ and $L_{d,p}$, the hole-rich region at $0 \le x < L_{p>n}$, and the electron-rich region at $L_{p>n} < x \le L_{\mathrm{MQW}}$ (= $L_{p>n} + L_{n>p}$). Electrons generated in $0 \le x < L_{p>n}$ have to travel within the hole-rich region by the distance of $L_{p>n} - x$ and hence, $\exp[-(L_{p>n} - x)/L_{d,n}]$ of them can be collected. The average collection efficiency of electrons directly generated in $0 \le x < L_{p>n}$, denoted by $\mathrm{CCE}_{n,\mathrm{direct}}$, becomes

$$\mathrm{CCE}_{n,\mathrm{direct}} = \frac{\int_0^{L_{p>n}} \exp[-\alpha x] \exp\left[-\left(L_{p>n} - x\right)/L_{d,n}\right] dx}{\int_0^{L_{p>n}} \exp[-\alpha x] dx}$$

$$= \frac{\alpha L_{d,n}}{1 - \alpha L_{d,n}} \frac{\exp\left[-\alpha L_{p>n}\right] - \exp\left[-L_{p>n}/L_{d,n}\right]}{1 - \exp\left[-\alpha L_{p>n}\right]} \tag{10a}$$

$$\approx \frac{L_{d,n}}{L_{p>n}} \left(1 - \exp\left[-\frac{L_{p>n}}{L_{d,n}}\right]\right) \tag{10b}$$

$$\approx \exp\left[-\frac{L_{p>n}}{2L_{d,n}}\right], \tag{10c}$$

where $\alpha$ is the absorption coefficient. Eq. (10b) assumes uniform photogeneration, which is acceptable in most MQW designs where $L_{\mathrm{MQW}} \le 2/\alpha$ for sufficient absorption. Eq. (10c) holds within the second-order approximation when $L_{p>n} \le 2L_{d,n}$, which should be satisfied for practical MQW cells. We can interpret Eq. (10c) as the average transport distance is a half of the minority-carrier region width when carriers are uniformly photogenerated inside the MQW.

Therefore, the total CCE for the uniform direct excitation including hole collection from $L_{p>n} < x \le L_{\mathrm{MQW}}$, denoted by $\mathrm{CCE}_{\mathrm{direct}}$, is given by

$$\mathrm{CCE}_{\mathrm{direct}} = \frac{L_{p>n}}{L_{\mathrm{MQW}}} \exp\left[-\frac{L_{p>n}}{2L_{d,n}}\right] + \frac{L_{n>p}}{L_{\mathrm{MQW}}} \exp\left[-\frac{L_{n>p}}{2L_{d,p}}\right]. \tag{11}$$

$\mathrm{CCE}_{\mathrm{direct}}$ has the maximum value given by

$$\max\left\{\mathrm{CCE}_{\mathrm{direct}}\right\} = \exp\left[-\frac{L_{\mathrm{MQW}}}{4\left\langle L_d \right\rangle_{p,n}}\right] \quad \text{when } \frac{L_{p>n}}{L_{d,n}} = \frac{L_{n>p}}{L_{d,p}}, \tag{12}$$

where $\left\langle L_d \right\rangle_{p,n} = \frac{1}{2}(L_{d,n} + L_{d,p})$ is the average drift length of electrons and holes. This explains why the p(p)n [Fig. 4(b)] and the MQW-top [Fig. 7(b)] samples under bias light, which have a better balance between $L_{p>n}$ and $L_{n>p}$ than the other samples, can efficiently collect carriers directly generated inside the MQW with wavelengths beyond 870 nm.

### *4.3. Output current under a given spectrum*

Based on the above calculation, the photocurrent $J_{ph}$, the total current $J$, the CCE (or IQE for negligible parasitic absorption) under any given illumination spectrum, and the EQE under monochromatic light can be written as

$$J_{ph} = q\left(G_{front}\,CCE_n + G_{back}\,CCE_p + G_{direct}\,CCE_{direct}\right) \tag{13}$$

$$J = J_{ph} - J_{dark}, \tag{14}$$

$$CCE = \frac{G_{front}\,CCE_n + G_{back}\,CCE_p + G_{direct}\,CCE_{direct}}{G_{front} + G_{back} + G_{direct}}, \tag{15}$$

$$EQE = \frac{G_{front}\,CCE_n + G_{back}\,CCE_p + G_{direct}\,CCE_{direct}}{N_\lambda} \tag{16a}$$

$$= A \times CCE, \tag{16b}$$

respectively, where $G_{front}$ and $G_{back}$ are the total generation rates in the device active regions above and below the MQW, which are the opposite for n-on-p structures, $G_{direct}$ is the total generation rate inside the MQW, $J_{dark}$ is the dark current, $N_\lambda$ is the illuminated photon flux, and $A$ is the absorptivity in the active region. Here, $G_{front}$, $G_{back}$, $G_{direct}$, and $N_\lambda$ can be obtained from the absorption coefficient and the incident light spectrum[2]. $CCE_n$, $CCE_p$, and $CCE_{direct}$ are given by Eqs. (8) and (11). This study was conducted at a constant voltage, but it should be noted that the CCE has a voltage dependency through $E$. In some cases, the carrier distribution, $L_{p>n}$ and $L_{n>p}$, may change at different voltages and affect the voltage dependency of CCE.

As a demonstration of the above formulae, we examined the CCE under a monochromatic light for the sample set with different MQW positions, which have a uniform field. The calculation result of Eq. (15) for this sample set is plotted in Fig. 7(c) using Eqs (8) and (11), the MQW physical parameters in Table 2, and the device parameters in Table 3. It agrees with the experimental result shown in Fig. 7(b), validating the proposed formulae for carrier collection. This model, described by a few simple expressions, confirms that the different carrier collection in identical MQWs placed in various positions is attributed to the carrier distribution, which cannot be easily interpreted by the conventional concept considering a single carrier in the potential field—the single-carrier picture.

[2]The standard solar spectrum for the single-junction application, the spectrum filtered by the upper subcells for the multi-junction application, or the monochromatic light for the operation under EQE measurement

**Table 3**

Device parameters of the sample set with different MQW positions for simulation in Fig. 7(c)

| Parameter | MQW-top | MQW-mid | MQW-bottom |
|---|---|---|---|
| p-emitter + top i-spacer[a] | 300 nm | 600 nm | 900 nm |
| n-base + bottom i-spacer[a] | 900 nm | 600 nm | 300 nm |
| Total MQW thickness ($L_{\text{MQW}}$) | | 300 nm | |
| Hole-rich region in MQWs ($L_{p>n}$)[b] | 140 nm | 0 nm | 0 nm |
| Electron-rich region in MQWs ($L_{n>p}$)[b] | 160 nm | 300 nm | 300 nm |
| Total i-region thickness ($L_i$) | | 1100 nm | |
| Applied voltage ($V$) | | 0.6 V | |
| Internal electric field ($E$)[c] | | 6.9 kV/cm | |
| Electron SRH lifetimes ($\tau_n$)[d] | | 300 ns | |
| Hole SRH lifetimes ($\tau_p$)[d] | | 35 ns | |

[a]These thicknesses were used to calculate the generation rates $G_{\text{front}}$, $G_{\text{back}}$, and $G_{\text{direct}}$. The Lambert-Beer law for each wavelength is applied.

[b]Obtained from the device simulation in Figs. 8(a)-(c). See appendix for the estimation of the carrier distribution profile without using a device simulator.

[c]Estimated from Eq. (A.5).

[d]Fitting parameters

For a rough estimation of the photocurrent, e.g. in the application of the device design, one may consider the special case in which the distribution is symmetric ($L_{p>n} = L_{n>p} = \frac{1}{2} L_{\text{MQW}}$). In this case, Eqs. (8) and (11) become

$$\text{CCE}_{n \text{ or } p} = \exp\left[ -\frac{L_{\text{MQW}}}{2L_d} \right], \tag{17}$$

$$\text{CCE}_{\text{direct}} = \left\langle \exp\left[ -\frac{L_{\text{MQW}}}{4L_d} \right] \right\rangle_{p,n}, \tag{18}$$

where $\langle \ \rangle_{p,n}$ is the symbol for the average between the electron and hole expressions.

The availability of analytical expressions gives a clear picture of the collection dynamics and the dependency on cell parameters. They provide cell design rules, which cannot be easily achievable by the numerical treatment using self-consistent simulators. For instance, Eq. (18) can be used for a design of MQW thickness $L_{\text{MQW}}$, and thus the number of well stacks. Figs. 9(a)-(b) illustrate the upper limit of $L_{\text{MQW}}$ to keep $\text{CCE}_{\text{direct}} \geq 0.9$ under the fixed electric field. From Eq. (18), these correspond to $L_{MQW}\big|_{\text{CCE}=0.9} = 0.42 L_{d,p} = 0.42 \mu_p \tau_p E$ for $\mu_n \tau_n = \mu_p \tau_p$ and $L_{MQW}\big|_{\text{CCE}=0.9} = 0.80 \mu_p \tau_p E$ for $\mu_n \tau_n = 10 \mu_p \tau_p$. In practice, the MQW total thickness $L_{\text{MQW}}$ cannot exceed the i-region thickness $L_i$, and the electric field $E$ decreases as $L_i$ increases. Figs. 9(c)-(d) similarly illustrate the upper limit of $L_{\text{MQW}}$ but under the fixed applied voltage $V$, assuming $L_{\text{MQW}} = L_i$ and thus $E = (V_{\text{bi}} - V)/L_{\text{MQW}}$ for the built-in voltage $V_{\text{bi}}$. That is, $L_{MQW}\big|_{\text{CCE}=0.9} = \sqrt{0.42 \mu_p \tau_p (V_{\text{bi}} - V)}$ for $\mu_n \tau_n = \mu_p \tau_p$ and $L_{MQW}\big|_{\text{CCE}=0.9} = \sqrt{0.80 \mu_p \tau_p (V_{\text{bi}} - V)}$ for $\mu_n \tau_n = 10 \mu_p \tau_p$.

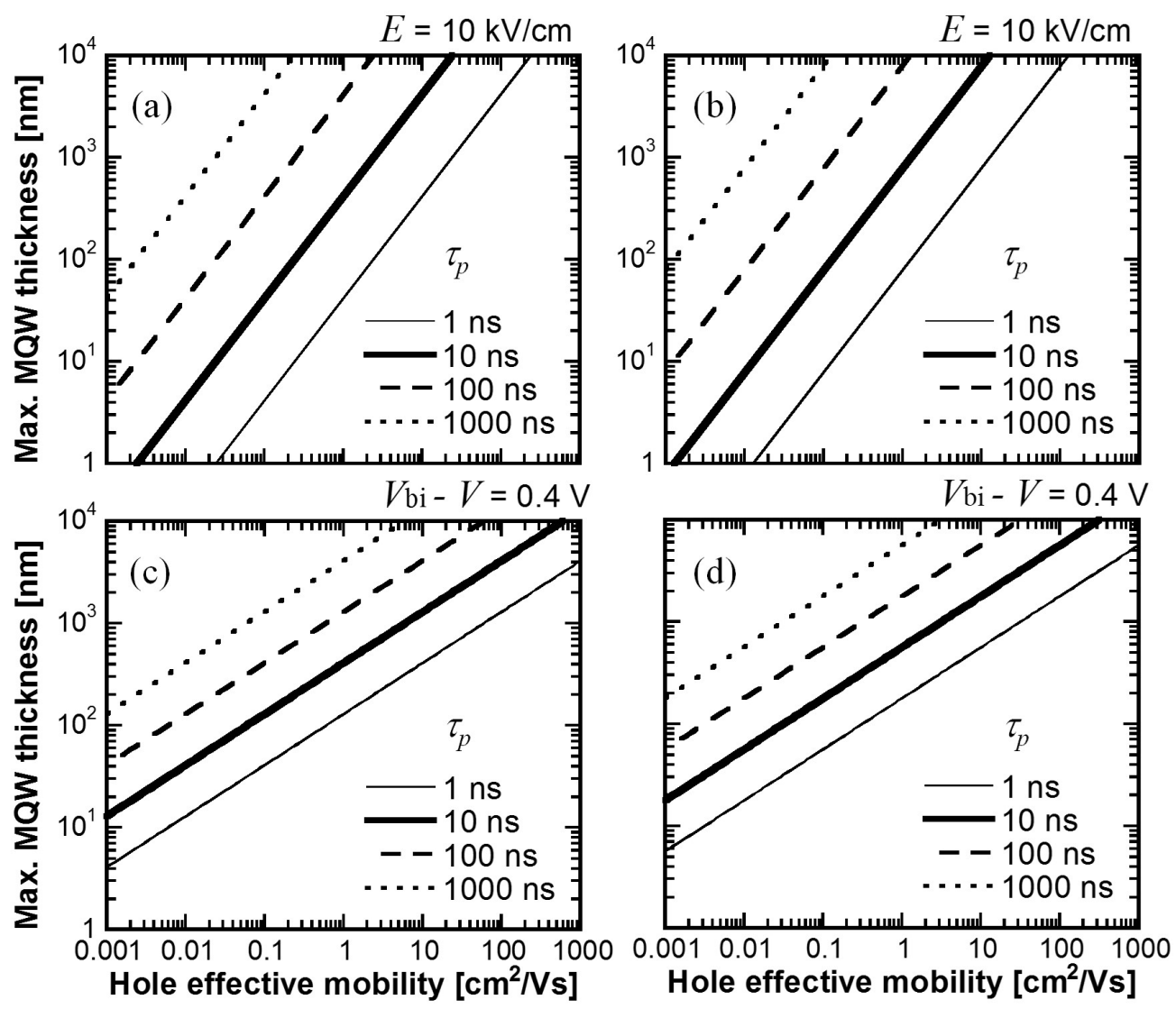

**Fig. 9.** Maximum MQW total thickness $L_{\mathrm{MQW}}$ to keep $\mathrm{CCE}_{\mathrm{direct}} \geq 0.9$ under various conditions. (a) Results under the same $\mu_{\mathrm{MQW}}\tau_{\mathrm{rec}}$ product for electrons and holes and (b) 10 times larger $\mu_{\mathrm{MQW}}\tau_{\mathrm{rec}}$ product for electrons than that for holes under the fixed electric field of 10 kV/cm. For other values of electric field, the maximum MQW thickness increases proportionally to the field. (c)-(d) Results under similar conditions but under the fixed applied voltage $V$ that satisfies $V_{\mathrm{bi}} - V = 0.4$ V, where the field is given by $E = (V_{\mathrm{bi}} - V)/L_{\mathrm{MQW}}$ .

Furthermore, as already mentioned, Eq. (12) implies that placing the MQW to balance the carrier distribution is an effective approach to suppress the recombination of carriers directly photogenerated in MQW and hence enhance the photocurrent. Nevertheless, we should keep in mind that the MQW position design in this way may simultaneously degrade the open-circuit voltage $V_{\mathrm{oc}}$, as commonly recognized that the SRH recombination of electrically injected carriers is significantly high at the position where $p \approx n$ . The way to deal with the $V_{\mathrm{oc}}$ trade-off depends on the application: it can be less problematic in multi-junction cell application in which the subcell photocurrent is more critical to the cell efficiency than the subcell voltage.

Even though only the SRH process is taken into account in this analysis, it can be extended to MQWs with more complicated recombination processes. The recombination rate in Eq. (7) $\mathrm{CCE}_n = \exp[\int dx/\mu_n \tau_n E\ ]$ can be simply replaced with the total rate $1/\tau_n = \sum 1/\tau_i$ of all recombination processes, and the following calculation can be carried out similarly. However, the final expression may not be as simple as when only the SRH process is dominant. For example, the radiative recombination lifetime of electrons, and in a similar way for holes, is given by $1/Bp(x)$ and has a complicated spatial distribution in the MQW. (See Appendix if the estimation of $p(x)$ and $\underline{n}(x)$ is needed.)

It should be noted that the analysis from Eq. (8) to (18) assumes the uniform electric field in the i-region. In devices with non-uniform electric field, e.g. with high background doping or under high concentration, where the charge accumulation in the MQW may screen the electric field [31,32], this assumption becomes invalid. Further extension of the analysis, using the general form of Eq. (7) instead of Eq. (8), is required to describe the operation of such MQW devices.

The above derivation can be applied not only to MQWs, but also to the carrier collection in other quantum structures and low-mobility absorbers inserted in the i-region of p-i-n junctions, such as dilute-nitride solar cells [33].

### 4.4. Relation with carrier escape time

It is sometimes more convenient to express the CCE using the escape time $\tau_{esc}$ and the recombination lifetime $\tau_{rec}$ as an analogy to the single-carrier picture. It has been derived in Ref. [24,25] that the effective carrier mobility in the MQW can be expressed by

$$\mu_{\mathrm{MQW}} = \left(1/\tau_{\mathrm{esc,f}} - 1/\tau_{\mathrm{esc,b}}\right) l/E \ , \tag{19}$$

where $1/\tau_{\mathrm{esc,f}}$ and $1/\tau_{\mathrm{esc,b}}$ are the rates for the forward escape to the next QW and the backward escape to the well behind, and $l$ is the MQW period. In most cases, $1/\tau_{\mathrm{esc,f}}$ and $1/\tau_{\mathrm{esc,b}}$ can be simply obtained from the carrier escape rate through a barrier with a decreased and an increased barrier height due to the electric field $E$, respectively, and their difference has a strong dependency on $E$. For example, the thermal process, whose escape rate through a potential barrier of effective height $V_b$ is given by $(1/\tau_{\mathrm{th0}})\exp\left[-qV_b/k_BT\right]$ [14], has an escape rate difference given by

$$\begin{aligned}\left(1/\tau_{\mathrm{esc,f}} - 1/\tau_{\mathrm{esc,b}}\right)_{\mathrm{thermal}} &= \left(1/\tau_{\mathrm{th0}}\right)\left(e^{-\frac{q\left(\overline{V}_b - wE/2\right)}{k_BT}} - e^{-\frac{q\left(\overline{V}_b + wE/2\right)}{k_BT}}\right) \\ &= \left(2/\tau_{\mathrm{th0}}\right) e^{-\frac{q\overline{V}_b}{k_BT}} \sinh\left(\frac{qwE}{2k_BT}\right),\end{aligned} \tag{20}$$

where $\overline{V}_b$ is the average barrier height, $k_BT$ is the thermal energy, $w$ is the well width, and $\tau_{\mathrm{th0}}$ is a constant.

By using Eq. (19) and the approximation $\exp\left[-Nx\right] \approx \left[1/\left(1+x\right)\right]^N$, the collection efficiency of electrons or holes in Eqs. (8) and (11) can be rewritten as

$$\mathrm{CCE} = \left[\frac{1/\tau_{\mathrm{esc,f}} - 1/\tau_{\mathrm{esc,b}}}{1/\tau_{\mathrm{esc,f}} - 1/\tau_{\mathrm{esc,b}} + 1/\tau_{\mathrm{rec}}}\right]^{\xi N_{\mathrm{minor}}} , \tag{21}$$

where $N_{\mathrm{minor}} = L_{\mathrm{minor}}/l$ is the number of wells in the minority-carrier region, and $\xi$ is 1 for carriers crossing MQW and $\frac{1}{2}$ for carriers uniformly generated inside the MQW. For the special case of symmetric carrier distribution, (21) becomes simpler using the total well number $N = L_{\mathrm{MQW}}/l$ :

$$\mathrm{CCE} = \left[ \frac{1/\tau_{\mathrm{esc,f}} - 1/\tau_{\mathrm{esc,b}}}{1/\tau_{\mathrm{esc,f}} - 1/\tau_{\mathrm{esc,b}} + 1/\tau_{\mathrm{rec}}} \right]^{\xi N/2} . \tag{22}$$

The expression is close to the simplified expression $\left[(1/\tau_{\mathrm{esc}})/(1/\tau_{\mathrm{esc}} + 1/\tau_{\mathrm{rec}})\right]^{N}$, but with much smaller number of wells to be considered. The *net* escape rate, including the backward escape process through $1/\tau_{\mathrm{esc}} = 1/\tau_{\mathrm{esc,f}} - 1/\tau_{\mathrm{esc,b}}$, should be considered particularly in low-field devices such as solar cells. Taking into account only the forward escape $1/\tau_{\mathrm{esc,f}}$ will largely overestimate the escape rate from MQWs; e.g., $1/\tau_{\mathrm{esc,f}}$ for the thermal escape in Eq. (20) is 5.7 times larger than the net escape rate $1/\tau_{\mathrm{esc}} = 1/\tau_{\mathrm{esc,f}} - 1/\tau_{\mathrm{esc,b}}$ for $w$ = 5 nm and $E$ = 10 kV/cm.

## 5. Conclusion

In this study, we investigated the collection mechanism of photogenerated carriers (photocurrent) in MQWs based on both experiment and simulation. By examining the MQW solar cells with different background doping levels, light biases, and positions, the results suggest that the concept of majority/minority carriers, even in low-doped or non-doped i-regions, is important in the collection process. Photogenerated electrons tend to recombine only in the hole-rich region, while holes tend to recombine only in the electron-rich region. This explains the efficient electron collection in the sample with $10^{15}$ $\mathrm{cm^{-3}}$-order n-type background doping even in the flat band, the recovery of carrier collection under bias light, and the different carrier collection behavior when the MQW is inserted in different positions in the i-regions. This behavior cannot be easily described by the single-carrier picture. The collective behavior of carriers, such as the carrier density and recombination profiles, is required for the understanding of device operation.

We proposed the model and derived the analytic expressions for the carrier collection efficiency taking into account such collective behavior, which can accurately predict the experimental results. As the impacts of cell parameters are expressed explicitly, it can be used as a design guide for optimizing MQW solar cells: for example, the required background doping, the proper MQW position, the optimal well number for sufficient absorption while maintaining efficient carrier collection, and so on. It is worth noting that our model is applicable to other types of devices containing quantum structures or low-mobility absorbers inside i-regions.

## Acknowledgments

This study was supported by the New Energy and Industrial Technology Development Organization (NEDO), Japan (P15003), and a Grant-in-Aid for JSPS Fellows (15J03447) from Japan Society for the Promotion of Science.

## Appendix: Rough estimation of carrier distribution

It is obvious from the above discussion that the estimation of carrier distribution is needed for calculating the CCE [Eqs. (8), (11), (21)]. As demonstrated in Figs. 3, 6, and 8, the density profiles can be obtained using a device simulator. However, it is more convenient for the device design if the carrier distribution profiles can be expressed explicitly. In this appendix, a rough estimation of the carrier distribution is discussed.

First, consider a p-i-n bulk solar cell with a uniform electric field $E$ (negligible background doping) and constant quasi-Fermi levels ($E_{Fn}$, $E_{Fp}$) across the i-region. Under the uniform field, the relative positions of the band edges from the quasi-Fermi levels [ $E_C(z)-E_{Fn}$ and $E_V(z)-E_{Fp}$, where $E_C$ and $E_V$ are the conduction and valence band edges] vary linearly with respect to position $z$ in the i-region. Hence, the carrier density is an exponential function of $z$ ( $0 \le z \le L_i$ ):

$$\log n(z) = \frac{L_i - z}{L_i}\log n(0) + \frac{z}{L_i}\log n(L_i), \tag{A.1a}$$

$$\log p(z) = \frac{L_i - z}{L_i}\log p(0) + \frac{z}{L_i}\log p(L_i). \tag{A.1b}$$

The boundary conditions can be given by

$$p(0) \equiv p_0 \approx N_A, \tag{A.2a}$$

$$n(L_i) \equiv n_{Li} \approx N_D, \tag{A.2b}$$

$$n(0) \equiv n_0 \approx n_i^2 \exp\left[qV/k_BT\right]/p_0, \tag{A.2c}$$

$$p(L_i) \equiv p_{Li} \approx n_i^2 \exp\left[qV/k_BT\right]/n_{Li}, \tag{A.2d}$$

where $z = 0$ and $z = L_i$ are the boundaries for the p-region (doping concentration $N_A$) and the n-region (doping concentration $N_D$), respectively, and $n_i$ is the intrinsic carrier density in the bulk.

When the MQW is inserted, electron and hole densities in the MQW region simply increase by factors of

$$\Xi_n = \frac{N_{c,\mathrm{MQW}}}{N_{c,\mathrm{bulk}}} e^{\frac{\Delta E_C}{k_BT}}, \tag{A.3a}$$

$$\Xi_p = \frac{N_{v,\mathrm{MQW}}}{N_{v,\mathrm{bulk}}} e^{\frac{\Delta E_V}{k_BT}}, \tag{A.3b}$$

respectively [34], where $N_{c,\mathrm{bulk}}$ and $N_{v,\mathrm{bulk}}$ are the effective densities of states in the conduction and valence bands of the surrounding bulk, $N_{c,\mathrm{MQW}}$ and $N_{v,\mathrm{MQW}}$ are the equivalent effective densities of states in the conduction and valence bands of the MQW, and $\Delta E_C$ and $\Delta E_V$ are the equivalent conduction and valence band offsets.

Furthermore, under bias light, the carrier density increment due to the electron photocurrent $\Delta J_n$ and hole photocurrent $\Delta J_p$ is given by Eq. (5). As a result, the electron and hole distribution profiles become

$$n(z)=\begin{cases} n_0^{\frac{L_i-z}{L_i}}\, n_{Li}^{\frac{z}{L_i}}+\dfrac{\Delta J_n(z)}{q\mu_{b,n}E} & ;\ \text{bulk} \\ \Xi_n n_0^{\frac{L_i-z}{L_i}}\, n_{Li}^{\frac{z}{L_i}}+\dfrac{\Delta J_n(z)}{q\mu_n E} & ;\ \text{MQW} \end{cases}, \tag{A.4a}$$

$$p(z)=\begin{cases} p_0^{\frac{L_i-z}{L_i}}\, p_{Li}^{\frac{z}{L_i}}+\dfrac{\Delta J_p(z)}{q\mu_{b,p}E} & ;\ \text{bulk} \\ \Xi_p p_0^{\frac{L_i-z}{L_i}}\, p_{Li}^{\frac{z}{L_i}}+\dfrac{\Delta J_p(z)}{q\mu_p E} & ;\ \text{MQW} \end{cases}, \tag{A.4b}$$

where $\mu_{b,n}$ and $\mu_{b,p}$ are the electron and hole mobilities in the surrounding bulk. Note that the quasi-Fermi levels, which are practically constant under a dark condition in most cells, are no longer uniform under illumination, namely Eq. (A.4). The electron field $E$ can be estimated from the built-in potential $V_{\text{bi}}$ and the total i-region thickness $L_i$:

$$E\approx\frac{V_{\text{bi}}-V}{L_i}. \tag{A.5}$$

As a further approximation, $\Delta J_n\approx q\left(G_{\text{front}}+\frac{1}{2}G_{\text{direct}}\right)$ and $\Delta J_p\approx q\left(G_{\text{back}}+\frac{1}{2}G_{\text{direct}}\right)$ are considered constant inside the MQW. (They are the opposite for n-on-p structures.) That is, $\Delta J_n$ and $\Delta J_p$ in the MQW can be roughly estimated with the average photocurrent in the MQW region without considering current drops due to recombination. Additionally, $\mu_{b,n}$ and $\mu_{b,p}$ are assumed to be sufficiently high. Then, the carrier densities become

$$n(z)\approx\begin{cases} n_0^{\frac{L_i-z}{L_i}}\, n_{Li}^{\frac{z}{L_i}} & ;\ \text{bulk} \\ \Xi_n n_0^{\frac{L_i-z}{L_i}}\, n_{Li}^{\frac{z}{L_i}}+\dfrac{G_{\text{front}}+\frac{1}{2}G_{\text{direct}}}{\mu_n E} & ;\ \text{MQW} \end{cases}, \tag{A.6a}$$

$$p(z)\approx\begin{cases} p_0^{\frac{L_i-z}{L_i}}\, p_{Li}^{\frac{z}{L_i}} & ;\ \text{bulk} \\ \Xi_p p_0^{\frac{L_i-z}{L_i}}\, p_{Li}^{\frac{z}{L_i}}+\dfrac{G_{\text{back}}+\frac{1}{2}G_{\text{direct}}}{\mu_p E} & ;\ \text{MQW} \end{cases}, \tag{A.6b}$$

This estimation may not be rigorous, but it is sufficient for approximately estimating the magnitude of the carrier densities. Fig. A.1(a) illustrates the estimated density profiles in the MQW-mid at 0.6 V under 1-sun bias light. The estimation using Eq. (A.6) provides the close similarity to the results in Fig. 8(b) obtained from the self-consistent simulator. The fractions of the electron-rich and hole-rich regions ($L_{n>p}/L_{\text{MQW}}$, $L_{p>n}/L_{\text{MQW}}$) in the MQW, inserted in different positions in the i-region estimated with Eq. (A.6) are plotted in Fig. A.1(b). The result confirms that the carrier distribution in the MQW is balanced when the MQW position is near, but not too close to, the p-emitter.

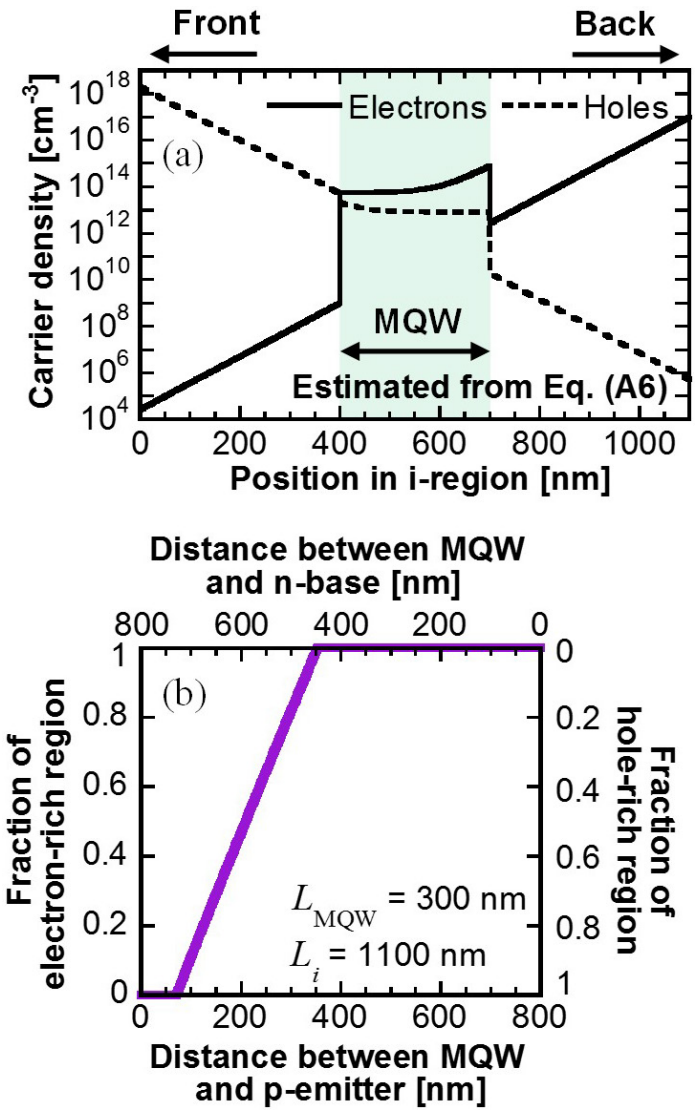

**Fig. A.1.** Estimation of carrier distribution profile in i-regions at 0.6 V using Eq. (A.6). (a) Electron and hole densities in the i-region of the MQW-mid under 1-sun bias light. (b) Fractions of electron-rich and hole-rich regions in the MQW with varied MQW positions. The MQW total thickness $L_{\mathrm{MQW}}$ was fixed at 300 nm and the i-region thickness $L_i$ was fixed at 1100 nm.

## References

[1] K. W. J. Barnham and G. Duggan, A new approach to high-efficiency multi-band-gap solar cells, J. Appl. Phys. 67 (1990) 3490-3493.

[2] K. Barnham, I. Ballard, J. Barnes, J. Connolly, P. Griffin, B. Kluftinger, J. Nelson, E. Tsui, and A. Zachariou, Quantum well solar cells, Appl. Surf. Sci. 113 (1997) 722-733.

[3] J. G. J. Adams, B. C. Browne, I. M. Ballard, J. P. Connolly, N. L. A. Chan, A. Ioannides, W. Elder, P. N. Stavrinou, K. W. J. Barnham, and N. J. Ekins-Daukes, Recent results for single-junction and tandem quantum well solar cells, Prog. Photovolt. Res. Appl. 19 (2011) 865-877.

[4] M. Kuramoto, H. Urabe, T. Nakano, A. Kawaharazuka, J. Nishinaga, T. Makimoto, and Y. Horikoshi, Optical properties of $Al_xGa_{1-x}As$/GaAs superlattice solar cells, J. Cryst. Growth 425 (2015) 333-336.

[5] B. Browne, J. Lacey, T. Tibbits, G. Bacchin, T. C. Wu, J. Q. Liu, X. Chen, V. Rees, J. Tsai, and J. G. Werthen, Triple-Junction quantum-well solar cells in commercial production, AIP Conf. Proc. 1556 (2013) 3-5.

[6] N. J. Ekins-Daukes, J. M. Barnes, K. W. J. Barnham, J. P. Connolly, M. Mazzer, J. C. Clark, R. Grey, G. Hill, M. A. Pate, and J. S. Roberts, Strained and strain-balanced quantum well devices for high-efficiency tandem solar cells, Sol. Energy Mater. Sol. Cells 68 (2001) 71-87.

[7] G. K. Vijaya, A. Alemu, and A. Freundlich, Dilute nitride multi-quantum well multi-junction design: a route to ultra-efficient photovoltaic devices, in Proceedings of SPIE - The International Society for Optical Engineering, San Francisco, 2011, 79330G.

[8] I. E. Hashem, C. Z. Carlin, B. G. Hagar, P. C. Colter, and S. M. Bedair, InGaP-based quantum well solar cells: growth, structural design, and photovoltaic properties, J. Appl. Phys. 119 (2016) 095706.

[9] K. Toprasertpong, H. Fujii, T. Thomas, M. Führer, D. Alonso-Álvarez, D. J. Farrell, K. Watanabe, Y. Okada, N. J. Ekins-Daukes, M. Sugiyama, and Y. Nakano, Absorption threshold extended to 1.15 eV using InGaAs/GaAsP quantum wells for over-50%-efficient lattice-matched quad-junction solar cells, Prog. Photovolt. Res. Appl. 24 (2016) 533-542.

[10] J. Nelson, M. Paxman, K. W. J. Barnham, J. S. Roberts, and C. Button, Steady-state carrier escape from single quantum wells, IEEE J. Quantum Electron. 29 (1993) 1460-1468.

[11] I. Serdiukova, C. Monier, M. F. Vilela, and A. Freundlich, Critical built-in electric field for an optimum carrier collection in multiquantum well p-i-n diodes, Appl. Phys. Lett. 74 (1999) 2812-2814.

[12] M. Elborg, T. Noda, and Y. Sakuma, Open-circuit voltage in AlGaAs solar cells with embedded GaNAs quantum wells of varying confinement depth, IEEE J. Photovoltaics 7 (2017) 162-168.

[13] M. Jo, Y. Ding, T. Noda, T. Mano, Y. Sakuma, K. Sakoda, L. Han, and H. Sakaki, Impacts of ambipolar carrier escape on current-voltage characteristics in a type-I quantum-well solar cell, Appl. Phys. Lett. 103 (2013) 061118.

[14] H. Schneider and K. v. Klitzing, Thermionic emission and Gaussian transport of holes in a GaAs/$Al_xGa_{1-x}As$ multiple-quantum-well structure, Phys. Rev. B 38 (1988) 6160-6165.

[15] A. M. Fox, D. A. B. Miller, G. Livescu, J. E. Cunningham, and W. Y. Jan, Quantum well carrier sweep out: relation to electroabsorption and exciton saturation, IEEE J. Quantum Electron. 27 (1991) 2281-2295.

[16] S. M. Ramey and R. Khoie, Modeling of multiple-quantum-well solar cells including capture, escape, and recombination of photoexcited carriers in quantum wells, IEEE Trans. Electron Devices 50 (2003) 1179-1188.

[17] G. K. Bradshaw, C. Z. Carlin, J. P. Samberg, N. A. El-Masry, P. C. Colter, and S. M. Bedair, Carrier transport and improved collection in thin-barrier InGaAs/GaAsP strained quantum well solar cells, IEEE J. Photovoltaics 3 (2013) 278-283.

[18] G. Zhou and P. Runge, Modeling of multiple-quantum-well p-i-n photodiodes, IEEE J. Quantum Electron. 50 (2014) 220-227.

[19] M. -J. Jeng, Y. -L. Lee, and L. -B. Chang, Temperature dependences of $In_xGa_{1-x}N$ multiple quantum well solar cells, J. Phys. D: Appl. Phys. 42 (2009) 105101.

[20] K. Driscoll, M. F. Bennett, S. J. Polly, D. V. Forbes, and S. M. Hubbard, Effect of quantum dot position and background doping on the performance of quantum dot enhanced GaAs solar cells, Appl. Phys. Lett. 104 (2014) 023119.

[21] H. Fujii, Y. Wang, K. Watanabe, M. Sugiyama, and Y. Nakano, Compensation doping in InGaAs / GaAsP multiple quantum well solar cells for efficient carrier transport and improved cell performance, J. Appl. Phys. 114 (2013) 103101.

[22] H. Fujii, K. Toprasertpong, K. Watanabe, M. Sugiyama, and Y. Nakano, Evaluation of carrier collection efficiency in multiple quantum well solar cells, IEEE J. Photovoltaics 4 (2014) 237.

[23] PVcell [Software]. Available from http://www.str-soft.com/products/solar/

[24] K. Toprasertpong, T. Inoue, K. Watanabe, T. Kita, M. Sugiyama, and Y. Nakano, Effective drift mobility approximation in multiple quantum-well solar cell, in Proceedings of SPIE - The International Society for Optical Engineering, 2016, 974315.

[25] K. Toprasertpong, S. M. Goodnick, Y. Nakano, and M. Sugiyama, Effective mobility for sequential carrier transport in multiple quantum well structures, arXiv:1706.05608.

[26] K. Toprasertpong, N. Kasamatsu, H. Fujii, T. Kada, S. Asahi, Y. Wang, K. Watanabe, M. Sugiyama, T. Kita, and Y. Nakano, Carrier time-of-flight measurement using a probe structure for direct evaluation of carrier transport in multiple quantum well solar cells, IEEE J. Photovoltaics 4 (2014) 1518-1525.

[27] K. Toprasertpong, T. Tanibuchi, H. Fujii, T. Kada, S. Asahi, K. Watanabe, M. Sugiyama, T. Kita, and Y. Nakano, Comparison of electron and hole mobilities in multiple quantum well solar cells using a time-of-flight technique, IEEE J. Photovoltaics 5 (2015) 1613-1620.

[28] Madelung, *Semiconductors: Data Handbook*, 3rd ed., Springer, Berlin, 2004.

[29] K. Toprasertpong, T. Inoue, A. Delamarre, K. Watanabe, J. -F. Guillemoles, M. Sugiyama, and Y. Nakano, Electroluminescence-based quality characterization of quantum wells for solar cell applications, J. Cryst. Growth 464 (2017) 94-99.

[30] E. D. Palik, *Handbook of optical constants of solids*, 1st ed., Academic Press, San Diego, 1997.

[31] O. Y. Raisky, W. B. Wang, R. R. Alfano, and C. L. Reynolds, Carrier screening effects in photoluminescence spectra of InGaAsP/InP multiple quantum well photovoltaic structures, Appl. Phys. Lett. 79 (2001) 430-432.

[32] W. Yanwachirakul, H. Fujii, K. Toprasertpong, K. Watanabe, M. Sugiyama, and Y. Nakano, Effect of barrier thickness on carrier transport inside multiple quantum well solar cells under high-concentration illumination, IEEE J. Photovoltaics 5 (2015) 846-853.

[33] D. J. Friedman, J. F. Geisz, S. R. Kurtz, and J. M. Olson, 1-eV solar cells with GaInNAs active layer, J. Cryst. Growth 195 (1998) 409-415.

[34] J. Nelson, I. Ballard, K. Barnham, J. P. Connolly, J. S. Roberts, and M. Pate, Effect of quantum well location on single quantum well p-i-n photodiode dark currents, J. Appl. Phys. 86 (1999) 5898-5905.